\documentclass[journal=jacsat,manuscript=article]{achemso}

\usepackage{chemformula} 
\usepackage[T1]{fontenc} 
\usepackage{threeparttable, tablefootnote}

\author{N. G. C. Brunken}

\affiliation[Leiden University]
{Leiden Observatory, 2300 RA Leiden, The Netherlands}

\email{brunken@strw.leidenuniv.nl}

\author{A. C. A.  Boogert}
\affiliation[Institute for Astronomy, University of Hawai’i at Manoa]{Institute for Astronomy, University of Hawai’i at Manoa 2680 Woodlawn Drive, Honolulu, HI 96822, USA}

\author{E. F. van Dishoeck}
\alsoaffiliation[Leiden University]{Leiden Observatory, 2300 RA Leiden, The Netherlands}
\affiliation[Max-Planck-Institut]{Max-Planck-Institut f\"{u}r Extraterrestrische Physik, Gie{\ss}enbachstra{\ss}e 1, 85748 Garching, Germany}

\author{N. J. Evans}
\affiliation[University of Texas]{Department of Astronomy, The University of Texas at Austin, 2515 Speedway, Stop C1400, Austin, Texas 78712-1205, USA}

\author{C. A. Poteet}
\affiliation[NV5 Geospatial Solutions]{NV5 Geospatial Solutions, Inc. 385 Interlocken Crescent, Suite 300 Broomfield, CO 80021, USA}

\author{K. Slavicinska}
\affiliation[Leiden University]{Leiden Observatory, 2300 RA Leiden, The Netherlands}

\author{L. Tychoniec}
\affiliation[Leiden University]{Leiden Observatory, 2300 RA Leiden, The Netherlands}

\author{P. Nazari}
\affiliation[eso]{European Southern Observatory, Karl-Schwarzschild-Strasse 2, 85748 Garching bei München, Germany 510}

\author{L. W. Looney}
\affiliation[university of illinois]{Department of Astronomy, University of Illinois, 1002 W. Green St., Urbana, IL, 61801}

\author{H. Tyagi}
\affiliation[tata institute]{Department of Astronomy and Astrophysics Tata Institute of Fundamental Research
4 Homi Bhabha Road, Colaba, Mumbai 400005, India}

\author{M. Narang}
\affiliation[tata institute]{Department of Astronomy and Astrophysics Tata Institute of Fundamental Research
4 Homi Bhabha Road, Colaba, Mumbai 400005, India}

\author{P. Klaassen}
\affiliation[Astronomy Technology Centre]{United Kingdom Astronomy Technology Centre, Edinburgh, Blackford Hill, Edinburgh, EH9 3HJ, United Kingdom}

\author{Y. Yang}
\affiliation[RIKEN]{RIKEN Cluster for Pioneering Research, Wakoshi, Saitama, 351-0106, Japan}

\author{P. J. Kavanagh}
\affiliation[Maynooth University]{Department of Experimental Physics, Maynooth University, Maynooth, Co. Kildare W23 F2H6, Ireland}

\author{S. T. Megeath}
\affiliation[University of Toledo]{Department of Physics and Astronomy, The University of Toledo, 2801 West Bancroft Street, Toledo, OH 43606, USA}

\author{M. E. Ressler}
\affiliation[JPL]{Jet Propulsion Laboratory, California Institute of Technology, 4800 Oak Grove Drive, Pasadena, CA 91109, USA}

\title {JWST observations of segregated $^{12}$CO$_2$ and $^{13}$CO$_2$ ices in protostellar envelopes}

\abbreviations{IR,NMR,UV}

\begin{document}



\begin{abstract}
The evolution of interstellar ices can be studied with thermal tracers such as the vibrational modes of CO$_2$ ice that show great diversity depending on their local chemical and thermal environment. Now with the wide spectral coverage and sensitivity of the James Webb Space Telescope we can obtain observations of the weak and strong CO$_2$ absorption features inhabiting the near- and mid-infrared spectral region. In this work we present observations of the 15.2 $\mu$m bending mode, the 4.39 $\mu$m stretching mode and the 2.70 $\mu$m combination mode of $^{12}$CO$_2$ and $^{13}$CO$_2$ ice in the high-mass protostar IRAS 20126 and the low-mass protostar Per-emb 35, two sources that show clear signs of ice heating.The 15.2 $\mu$m bending mode of both protostars shows the characteristic double peak profile that is associated with pure CO$_2$ ice and a sharp short-wavelength peak is observed at 4.38 $\mu$m in the $^{13}$CO$_2$ bands of the two sources. Furthermore, a narrow short-wavelength feature is detected at 2.69 $\mu$m in the $^{12}$CO$_2$ combination mode of Per-emb 35.
We perform a consistent profile decomposition on all three vibrational modes and show that the profiles of all three bands can be reproduced with the same linear combination of CO$_2$ ice in mixtures with mostly CH$_3$OH and H$_2$O ices when the ices undergo segregation due to heating. The findings show that upon heating, CO$_2$ ice is likely segregating from mostly the water-rich ice layer and the CO$_2$-CH$_3$OH component becomes dominant in all three vibrational modes. Additionally, we find that the contribution of the different CO$_2$ components to the total absorption band is similar for both $^{12}$CO$_2$ and $^{13}$CO$_2$. This indicates that fractionation processes must not play a significant role during the different formation epochs, H$_2$O-dominated and CO-dominated, of the CO$_2$ ices and that ratio persists through the heating stage. 
We quantify the $^{12}$CO$_2$ and $^{13}$CO$_2$ ice column  densities and derive $^{12}$C/$^{13}$C$_{ice}$ = 90 $\pm$ 9 in IRAS 20126, a value that is lower compared to what was previously reported for warm gaseous CO in this source. Finally, we report the detection of the $^{13}$CO$_2$ bending mode of pure CO$_2$ ice at 15.64 $\mu$m in both IRAS 20126 and Per-emb 35. 

\end{abstract}

\section{Introduction}

The formation of an infant star in a collapsing dark molecular cloud marks the beginning of the protostellar stage. These stellar cradles are also the formation sites of interstellar ices where the low temperatures and higher densities enable atoms and small molecules to stick to the surfaces of cold dust grains. These small species will subsequently react to create the molecules that later evolve and become the prebiotic material that is incorporated into planets \citep{dishoeck2014}. Consequently, it is imperative to study the chemical journey of interstellar ices in order to define the initial conditions that could potentially lead to habitability. 
One process that can significantly alter the structure and composition of interstellar ices is heating by the central protostar \citep{boogert2015observations,cuppen2024,visser2009chemical, oberg2011spitzer,boogert2008}. This thermal processing also alters the infrared absorption features of these ices, a well-known example being the crystalline profile of water ice \citep{fraser2001}, and these ice absorption bands can therefore act as probes when examining physicochemical processes. 

The vibrational modes of CO$_2$ ice in particular have been well studied for their ability to trace ice heating and composition \citep{ehrenfreund1997infrared,ehrenfreund1998ice,ehrenfreund1999laboratory, gerakines1999infrared,boogert1999iso,pontoppidan2003m,pontoppidan2008c2d,brunken2024a}. Studies have shown that both the 15.2 $\mu$m bending mode of $^{12}$CO$_2$ \citep{gerakines1999infrared,boogert1999iso,pontoppidan2008c2d,poteet2013anomalous,isokoski2013highly, brunken2024a} and the 4.39 $\mu$m stretching mode of $^{13}$CO$_2$ \citep{boogert1999iso, brunken2024a} change dramatically depending on the line of sight. The 15.2 $\mu$m bending mode, for instance, is known to split into two peaks and the appearance of a second peak is observed at 4.38 $\mu$m in infrared spectra of luminous protostars. Both spectral features have been attributed to segregated CO$_2$ ice, a process in which CO$_2$ molecules cluster together and form inclusions of pure CO$_2$ ice in the otherwise mixed ice mantles upon protostellar heating \citep{ehrenfreund1998ice, pontoppidan2008c2d,oberg2011spitzer}.

The era of the James Webb Space Telescope (JWST) provides new and unique opportunities to study the vibrational modes of CO$_2$ ice at higher S/N in high-mass and solar-mass protostars. In particular, its exceptional sensitivity enables observations of weak features such as the 2.70 $\mu$m $^{12}$CO$_2$ combination mode and the 4.39 $\mu$m $^{13}$CO$_2$ stretching mode. These weaker bands have the additional advantage of being unsusceptible to the grain shape and size effects that can further alter the band profiles \citep{ehrenfreund1996,dartois2006spectroscopic, dartois2022influence}. Finally, the wide spectral coverage of the JWST allows access to the strong $^{12}$CO$_2$  4.27 $\mu$m stretching mode and 15.2 $\mu$m bending mode \citep{brunken2024a,brunken2024b} enabling a complete study of all the CO$_2$ ice absorption features. 

In this work we use JWST observations to investigate the environment of CO$_2$ ice in the low-mass-protostar Per-emb 35 and the high mass protostar IRAS 20126+4104 (hereafter IRAS 20126) as part of the JWST Observations of Young protoStars (JOYS+) program \citep{dishoeck2023,beuther2023,dishoeck2025} and the Investigating Protostellar Accretion Across the Mass Spectrum (IPA) program \citep{federman2023, narang2023,rubinstein2023}. Both sources show spectral signatures of thermally processed ices. We perform for the first time a consistent profile decomposition of three $^{12}$CO$_2$ bands: the $^{12}$CO$_2$ $\nu_2$ bending mode (15.2 $\mu$m), the $^{13}$CO$_2$ $\nu_3$ stretching mode (4.39 $\mu$m) and the $\nu_1$ + $\nu_3$ combination mode (2.70 $\mu$m).
This paper is structured as follows. In Sect. 2 we present our observations and provide the methods used for the profile analysis. In Sect. 3 we present the spectral decompositions. The results are discussed in Sect. 4, and the main points of this work are summarized in Sect. 5.

\section{Observations and Methods}
\subsection{Observations}
\label{sec:observations}

\begin{figure}[h!]
    \centering
     \includegraphics[width=1\hsize]{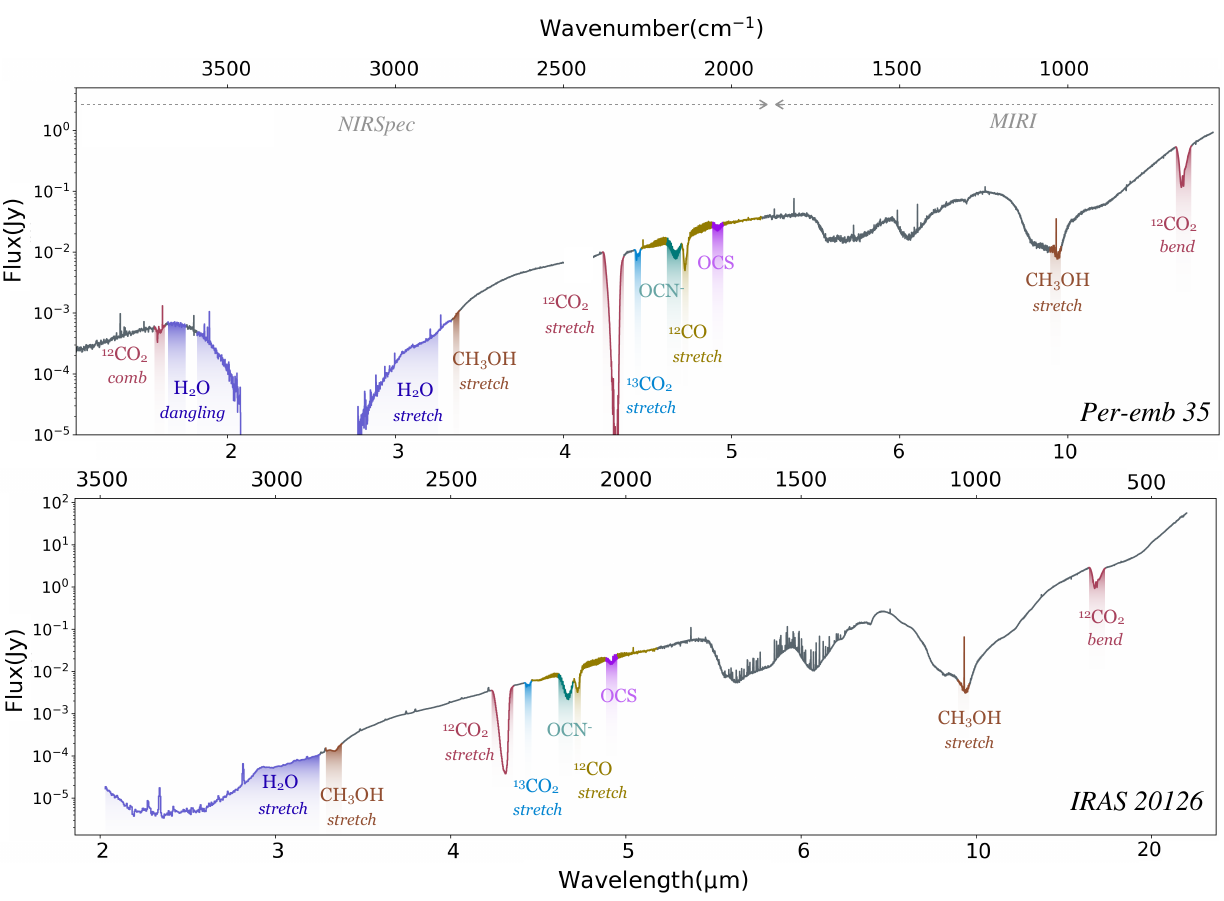}
    \caption{Full JWST NIRSpec and MIRI spectra for Per-emb 35 and IRAS 20126. The ice absorption features are shaded in color. }
    \label{fig:full-per35}
\end{figure}

The observations of Per-emb 35 were taken as part of the JOYS+ Cycle 1 NIRSpec program (PI: E.F. van Dishoeck, ID: 1960) and MIRI (PI: M. E. Ressler, ID: 1236) programs. The data consist of NIRSpec (1 - 5 $\mu$m) spectra observed using the G235H and G395H modes ($R$ = $\lambda$ /$\Delta$ $\lambda$ = 2700) and MIRI (5 - 28 $\mu$m) spectra observed using MIRI MRS gratings (A, B, and C). The IPA observations of IRAS 20126 were obtained as part of the IPA Cycle 1 GO program (PI: T.Megeath, ID: 1802) using the NIRSpec G395M mode ($R$ = $\lambda$ /$\Delta$ $\lambda$ = 1000) and MIRI MRS gratings. Both spectrometers are integral field units and the spectra were extracted centered on the infrared source using a 3$\lambda$/D cone aperture. We note that a circle aperture with  fixed radius provides similar spectra as a cone aperture extraction. Absolute calibration errors are less than 5\% for both instruments. Further observational details, data reduction methods, information on the spectral extractions coordinates and source properties are provided Table \ref{Tab:properties} and in references \citep{brunken2024a,brunken2024b,gelder2024,federman2023,rubinstein2023,narang2024, gelder2024b}. An overview of the ice bands discussed in this work is presented in Figure \ref{fig:full-per35} for Per-emb 35 and IRAS 20126.

\subsection{Continuum subtraction}

For the continuum subtractions on the 2.70 $\mu$m band of Per-emb 35 and the 4.39 $\mu$m bands of Per-emb 35 and IRAS 20126, we first applied a local continuum over the ice bands and subsequently used this to convert the spectra from flux scale to optical depth scale using equation (\ref{eq:OD}),

\begin{equation}
\label{eq:OD}
    \tau^{obs}_{\lambda} = -ln\biggl(\frac{F^{obs}_{\lambda}}{F^{cont}_{\lambda}}\biggl),
\end{equation}
where $F^{obs}_{\lambda}$ is the observed flux and $F^{cont}_{\lambda}$ is the flux of the continuum. Further details on the continuum placement and the anchor points used in each spectral region are described with supporting figures in \citet{brunken2024b}.

The continuum in the 15 $\mu$m region was subtracted by first fitting a global continuum following the methods presented in \citet{boogert2008,poteet2011}. After subtracting the continuum and converting the spectra to optical depth scale, we fitted the silicate template of GCS 3 to the 9.7 $\mu$m and 18 $\mu$m silicate features and subtracted this from the spectra \citep{kemper2004}. The line of sight of GCS 3 passes through a large column density of a diffuse cloud containing silicates but little ices. Following the silicate subtraction we fitted laboratory data of water ice at various temperatures \citep{slavicinska2024} to subtract the water libration mode at 12.6 $\mu$m from the spectrum of Per-emb 35. The continuum subtraction and water libration mode subtraction for Per-emb 35 are presented in Figures \ref{fig:continuum-per35} and \ref{fig:water-lib-per35} in the Supporting Information section.

The spectrum of IRAS 20126 proved to be more difficult to fit with the water laboratory data currently available to us and we opted instead to fit a local continuum over the CO$_2$ 15.2 $\mu$m band to simulate the wing of the water libration mode and subtracted this from the spectrum. We note that this method of simulating the overlapping spectral features in this region with local continua and Gaussian curves is similar to the methods applied in \citet{gerakines1999infrared} and \citet{pontoppidan2008c2d}. The final spectra of IRAS 20126 are shown on optical depth scale in Figure \ref{fig:continuum-iras20126}. The uncertainty on the continuum placement can account for up to 20\% in the error budget \citep{brunken2024b}. 

\subsection{Spectral decomposition}

The spectral decompositions of the  2.70 $\mu$m, the 4.39 $\mu$m and the 15.2 $\mu$m bands of CO$_2$ were performed using laboratory spectra. The laboratory data \citep{ehrenfreund1997infrared, ehrenfreund1999laboratory,van2006infrared} that provided the best fit are presented in Table \ref{Tab:spectra} and are publicly available on the Leiden Ice Data Base (LIDA) \citep{rocha2022lida}.

\begin{center}
\begin{table*}[hbt!]
\caption{Laboratory spectra.}
\small
\centering
\begin{tabular}{lcccl}
\hline \hline
Ice sample & Ratio & $T$(K) & Resolution (cm$^{-1}$) & Reference   \\         
\hline 

CO$_2$:H$_2$O & 1:1 & 100 & 1 & \citet{ehrenfreund1999laboratory}  \\

CO$_2$:CH$_3$OH  & 1:1 & 115 & 1 & \citet{ehrenfreund1999laboratory} \\

CO$_2$:CO & 1:1 & 15 & 0.5 &  \citet{van2006infrared} \\

CO$_2$:CO & 1:2 & 25 & 0.5 &  \citet{van2006infrared} \\

CO$_2$ & Pure & 80 & 1 & \citet{ehrenfreund1997infrared} \\

\hline
\end{tabular}
\label{Tab:spectra}
\end{table*}  
\end{center}

To correct for the grain shape and size effects that affect the strong vibrational mode at 15.2 $\mu$m \citep{dartois2022influence}, we used {\tt optool} \citep{dominik2021} and performed corrections for a  continuous distribution of ellipsoids (CDE) on the laboratory spectrum of pure CO$_2$. These effects are negligible for the weaker vibrational mode at 2.70 $\mu$m and the 4.39 $\mu$m band of $^{13}$CO$_2$ because the isotopologue is diluted in $^{12}$CO$_2$. The particle shape effects are also negligible for the laboratory spectra where CO$_2$ is diluted in other species \citep{tielens1991interstellar, ehrenfreund1997infrared}.

The bands were fitted using a $\chi^2$ minimization routine that provides the best linear combination of five components each representative of CO$_2$ in a specific chemical environment \citep{pontoppidan2008c2d}. The best fit was selected based on the lowest $\chi^2$ value and for the components we used laboratory spectra of the binary ices CO$_2$:H$_2$O, CO$_2$:CH$_3$OH, CO$_2$:CO and pure CO$_2$ ice. Given the number of mixing ratios available for each binary ice, we opted to use the mixing ratios determined in previous CO$_2$ studies \citep{pontoppidan2008c2d,brunken2024a} as a starting point and introduced new mixing ratios when necessary. As input spectra for the $\chi^2$ minimization routine, we used all available temperature measurements for a given mixing ratio.

The 15.2 $\mu$m band of IRAS 20126 was first fitted with the linear combination used in \citet{brunken2024a} to fit the $^{13}$CO$_2$ 4.39 $\mu$m  band in this source. If this initial linear combination failed to reproduce the 15.2 $\mu$m band, the 4.39 $\mu$m band analysis was re-fitted with a combination of different laboratory spectra. For example, a new mixing ratio could be introduced before running the $\chi^2$ routine to find a new best fit for the 4.39 $\mu$m feature. The 4.39 $\mu$m band analysis was re-visited because the components are better isolated and therefore more distinguishable in this vibrational mode compared to the 15.2 $\mu$m bending mode. 

Once a new best fit was found for the 4.39 $\mu$m band, we ran the $\chi^2$ routine with all the available temperatures of this new mixing ratio on the 15.2 $\mu$m band to test whether the routine would select the same spectra to fit this feature. If the band was instead fitted with a different linear combination, we used the selected five spectra to fit the at 4.39 $\mu$m and test if this combination could also reproduce the $^{13}$CO$_2$ feature. If a spectrum in this combination failed to reproduce the 4.39 $\mu$m band, it was removed from the list of input spectra, and we ran the $\chi^2$ routine again on the 15 $\mu$m band with the remaining input spectra. This process was repeated until consistent results were found between the bands, after which the 2.70 $\mu$m band was included in the analysis to expand the consistency check. The routine provides 1$\sigma$ uncertainties on each fitted component. This is a margin for how much the contribution of each component can be increased or decreased before the band profile is no longer reproduced by the linear combination. This multi-band analysis allowed us to provide additional constraints on the components and lower some of the degeneracies between the laboratory spectra. Further details on the fitting of our sources are presented in the Analysis section.

\subsection{Column densities}

The column densities are calculated using equation \ref{eq:CD}:

\begin{equation}
\label{eq:CD}
    N = \frac{\int\tau_\nu d\nu}{A},
\end{equation}

where $\int$$\tau_\nu$ d$\nu$ is the integrated optical depths under the absorption feature, and $A$ is the corresponding band strength of the vibrational mode. For the band strength we used the values determined by \citet{gerakines1995} and corrected by \citet{bouilloud2015}. The corrected band strengths used in this work are presented Table \ref{Tab:bandstrengths1}. In order to facilitate comparison with studies that do not use the corrected band strengths, we have included the correction factors that can be used to convert the column densities. The column densities derived in this work need to be multiplied with these correction factors in order to compare them with column densities derived using the uncorrected band strengths.

\begin{center}
\begin{table*}[hbt!]
\caption{Band strengths of CO$_2$ ice.}
\small
\centering
\begin{tabular}{lccc}
\hline \hline
Position ($\mu$m  & $A$ (cm\, molecule$^{-1}$)  & Correction Factor & References
\\     
\hline 

2.70 &  2.1 $\times$ $10^{-18}$ & 1.4 &  \citet{gerakines1995} (corrected) \\

4.27 &  1.1 $\times$ $10^{-16}$ & 1.45 & \citet{gerakines1995} (corrected) \\

15.2 &  1.6 $\times$ $10^{-17}$  & 1.45 & \citet{gerakines1995} (corrected) \\

4.39 &  1.15 $\times$ $10^{-16}$  & 1.47 & \citet{gerakines1995} (corrected) \\
\hline

\hline
\end{tabular}
\label{Tab:bandstrengths1}
\begin{tablenotes}\footnotesize
\item{\textbf{Notes.} The corrected values of \citet{gerakines1995, gerakines2005} were taken from \citet{bouilloud2015}. The column densities derived in this work can be multiplied by the correction factors to compare them with the values derived using the band strengths reported in \citet{gerakines1995} and \citet{gerakines2005}. } 
\end{tablenotes}
\end{table*}  
\end{center}

\section{Analysis}
\label{sec:analysis}

In the following sections we present the profile analysis of the $^{12}$CO$_2$  $\nu_2$ bending mode (15.2 $\mu$m), the $^{13}$CO$_2$ $\nu_3$ stretching mode (4.39 $\mu$m) and the $^{12}$CO$_2$ $\nu_1$ + $\nu_3$ combination mode (2.70 $\mu$m). The strong $^{12}$CO$_2$ $\nu_3$ stretching mode at 4.27 $\mu$m is observed in both sources but is saturated ($\tau$ > 5). Its true optical depth therefore remains uncertain. Furthermore, its profile is highly sensitive to grain shape and size effects \citep{dartois2006spectroscopic, dartois2022influence}. Consequently, we opted to exclude this band from this analysis. We also report the detection of a weak feature at 15.64 $\mu$m which we assign to the bending mode of $^{13}$CO$_2$ ice. 

\subsection{High-mass protostar: IRAS 20126}

\begin{figure}[h!]
    \centering
     \includegraphics[width=1\hsize]{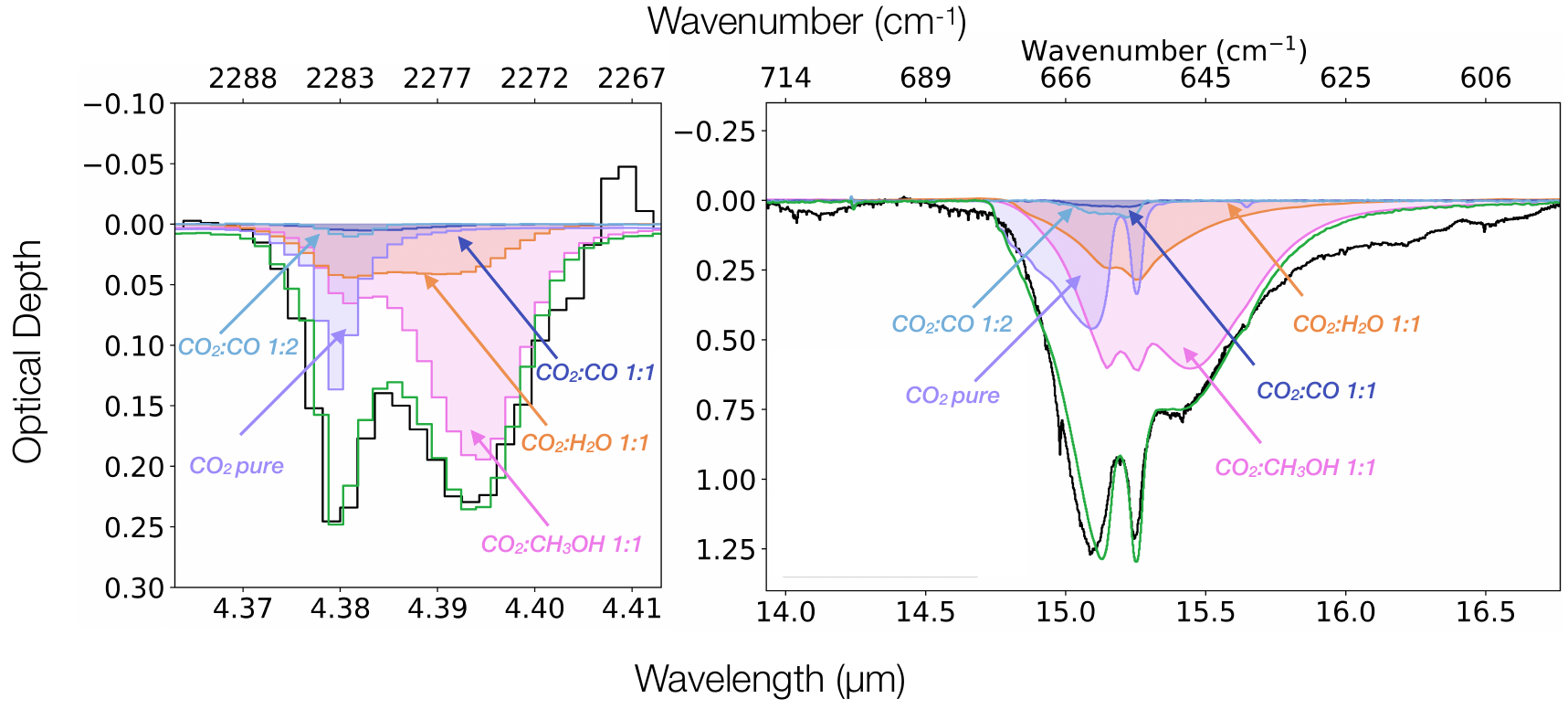}
    \caption{Band profile analysis of IRAS 20126. Left: Decomposition of the 4.39 $\mu$m $^{13}$CO$_2$ band. Right: Decomposition of the 15.2 $\mu$m $^{12}$CO$_2$ band. The purple, pink, orange, light blue and dark blue shaded areas correspond to the pure CO$_2$ 80 K, CO$_2$:CH$_3$OH 1:1 115 K, CO$_2$:H$_2$O 1:1 100 K, CO$_2$:CO 1:2 25 K and CO$_2$:CO 1:1 15 K component, respectively. Finally the green line shows the sum of all the components. The poor fit at 16.2 $\mu$m and 16.5 $\mu$m is likely due to absorption features of crystalline silicates.}
    \label{fig:an-IRAS20126}
\end{figure}

In Figure \ref{fig:an-IRAS20126} we present the profile analysis of the 4.39 $\mu$m band of $^{13}$CO$_2$ and the 15.2 $\mu$m band of $^{12}$CO$_2$ in the high-mass source IRAS 20126. Currently there are no NIRSpec G235M observations of this source and the 2.70 $\mu$m combination mode is therefore not available.

\begin{center}
\begin{table*}[hbt!]
\caption{Fraction of integrated optical depth IRAS 20126}
\small
\centering
\begin{tabular}{lccc}
\hline \hline
Component  & Mixture & 4.39 $\mu$m (\%) & 15.2 $\mu$m  (\%)
\\     
\hline 

CO$_2$:H$_2$O & 1:1 & 20 $\pm$ 10   & 18 $\pm$ 1   \\
CO$_2$:CH$_3$OH & 1:1 & 60 $\pm$ 2  & 58 $\pm$ 1   \\
CO$_2$:CO& 1:1 & 1  $\pm$ $^{+6}_{-1}$ & 	2 $\pm$ 1   \\
CO$_2$:CO &1:2  &  1 $\pm$ $^{+2}_{-1}$ & 	2 $\pm$ 1 \\
CO$_2$ & Pure	& 18  $\pm$ 2 & 21  $\pm$ 1   \\
\hline

\end{tabular}
\label{Tab:IRAS20126}
\end{table*}  
\end{center}

The $^{13}$CO$2$ feature of IRAS 20126 was first modeled by \citet{brunken2024a}, who performed a phenomenological decomposition of the ice feature and identified the peak positions and widths of three main components: a short-wavelength peak, a long-wavelength peak and a middle component (Supplementary Table \ref{Tab:gaus}). The band was subsequently fitted with a linear combination of five laboratory spectra based on the study by \citet{pontoppidan2008c2d} of the 15.2 $\mu$m bending mode. The components are representative of CO$_2$ ice in the following environments: pure CO$_2$ ice, CO$_2$ in an H$_2$O-rich environment, CO$_2$ mixed with CH$_3$OH, CO$_2$ diluted in CO and CO$_2$ mixed with CO in equal ratios.
\citet{brunken2024a} compared the width and central positions of the short-wavelength, long-wavelength and middle component of the 4.39 $\mu$m band (Supplementary Table \ref{Tab:gaus}) with those of various laboratory spectra and selected five laboratory spectra to fit this absorption band: CO$_2$:H$_2$O 1:10 10 K, CO$_2$:CH$_3$OH 1:10 10 K, CO$_2$:CO 1:1 15 K, CO$_2$:CO 1:2 25 K and pure CO$_2$ 80 K. For further details on this selection process, we refer the reader to \citet{brunken2024a}.

\begin{figure}[h!]
    \centering
     \includegraphics[width=1\hsize]{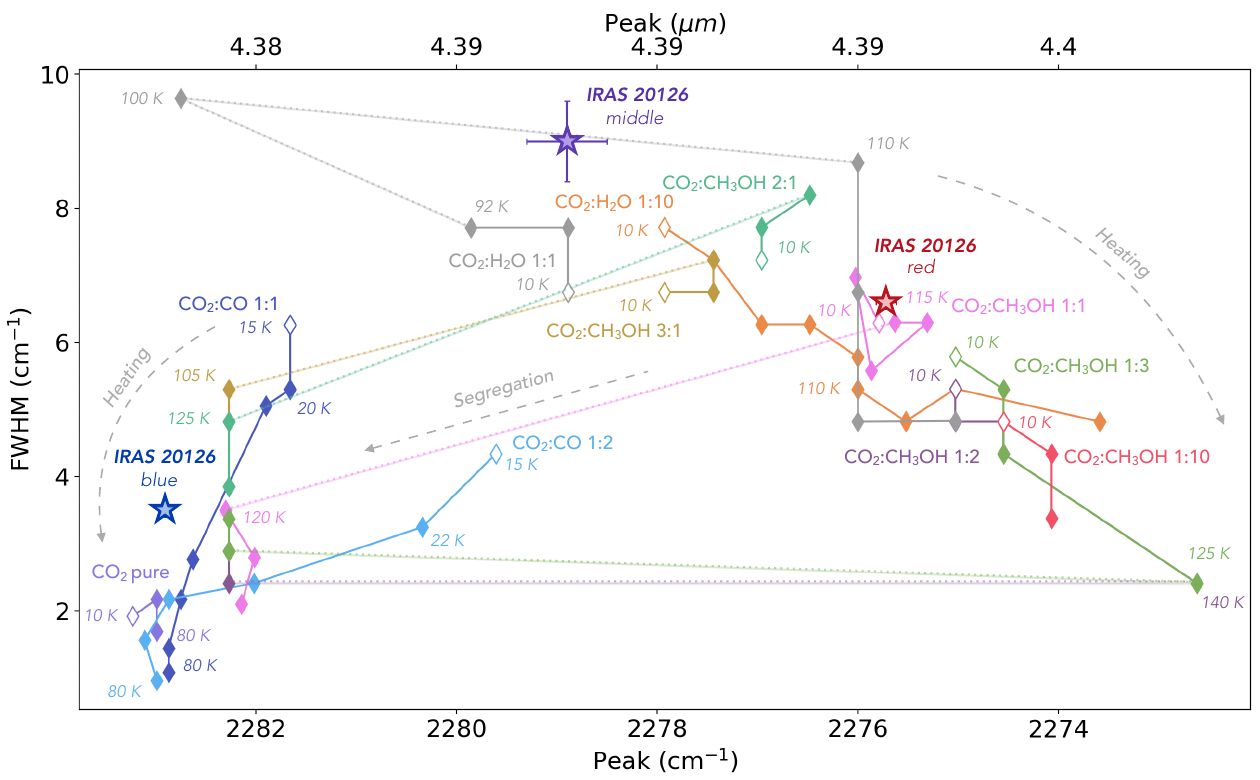}
    \caption{Full width at half maximum and peak positions of the available laboratory CO$_2$ spectra of the 4.39 $\mu$m $^{13}$CO$_2$ band. The arrows indicate how ice heating and segregation is affecting these quantities. The width and central positions measured for the long-wavelength, short-wavelength and middle components of the $^{13}$CO$_2$ band are shown as colored stars. }
    \label{fig:boogert}
\end{figure}

In Figure \ref{fig:boogert} we present an extended version of Figure 3 from \citet{brunken2024a} showing the evolution of the full width at half maximum (FHWM) and central positions of the available CO$_2$ laboratory spectra for the 4.39 $\mu$m band of $^{13}$CO$_2$. The widths and peak positions of the long-wavelength, short-wavelength and middle components measured for the $^{13}$CO$_2$ band in IRAS 20126 are also shown. 


The 15.2 $\mu$m bending mode of IRAS 20126 was initially modeled using the same combination of laboratory spectra selected by \citet{brunken2024a} as a starting point. Given the susceptibility of the 15.2 $\mu$m band to grain shape and size effects, we applied a CDE correction on the laboratory spectrum of pure CO$_2$ prior to the fitting. This five-component linear combination failed to reproduce the 15.2 $\mu$m absorption band however because the laboratory spectrum of CO$_2$:CH$_3$OH 1:10 at 10 K is too red-shifted to fit the shoulder located at 15.4 $\mu$m. Laboratory spectra of CO$_2$:CH$_3$OH 1:10 ices at higher temperatures also failed to reproduce this shoulder. \citet{brunken2024a} also provided an alternative three-component linear combination of solely heated ices but this failed to reproduce the 15.2 $\mu$m feature as well.  

Consequently, we revisited the profile analysis of the $^{13}$CO$_2$ 4.39 $\mu$m band. The width and central position of the long-wavelength feature observed in the $^{13}$CO$_2$ band (Table \ref{Tab:gaus}) were compared to laboratory spectra of CO$_2$:CH$_3$OH ices and the results indicate that the 1:1 mixing ratio produces bands with properties that coincide with this peak (Figure \ref{fig:boogert}). The CO$_2$:CH$_3$OH 1:10 laboratory spectra were then exchanged in favor of the 1:1 spectra and we ran the $\chi^2$ fitting routine with these new input spectra as described in Section 2. Our findings show that the band profiles of both vibrational modes are successfully reproduced with the CO$_2$:CH$_3$OH 1:1 spectrum at 115 K as shown in Figure \ref{fig:an-IRAS20126}. 

While the CO$_2$:H$_2$O 1:10 at 10 K successfully fitted both vibrational modes, the possibility of a warm CO$_2$:H$_2$O component was also considered given that both the 15.4 $\mu$m shoulder and the 4.39 $\mu$m peak are better fitted with high-temperature CO$_2$:CH$_3$OH spectra. Ice heating causes the CO$_2$:H$_2$O 1:10 spectra to shift to significantly longer wavelengths however (Figure \ref{fig:boogert}) and these spectra no longer fit the $^{13}$CO$_2$ band in a five-component linear combination. Therefore, we opted to run the $\chi^2$ routine with spectra of CO$_2$:H$_2$O ices in different mixing ratios. Given the limited available laboratory data, we selected the CO$_2$:H$_2$O 1:1 mixture; the abundance of CO$_2$ with respect to H$_2$O is $\sim$ 20 - 50 \% in the protostellar envelopes \citep{boogert2015observations}. 

Figure \ref{fig:an-IRAS20126} shows that the band profiles of both the $^{13}$CO$_2$ 4.39 $\mu$m and the $^{12}$CO$_2$ 15.2 $\mu$m bands are successfully fitted with high-temperature spectra of this 1:1 mixing ratio. The Supplementary Figure \ref{fig:an-IRAS20126-cold} shows the alternative fit with the cold CO$_2$:H$_2$O 1:10 spectrum at 10 K. Thus, there is a clear degeneracy between these CO$_2$:H$_2$O laboratory spectra. In the Discussion section we will argue why the high-temperature water spectra are better suited for the analysis of the vibrational modes in IRAS 20126 and Per-emb 35. 

In Table \ref{Tab:IRAS20126} we present the fraction of each component with respect to the total integrated optical depth including their 1$\sigma$ uncertainties. The fraction of integrated optical depth is consistent between the two bands within these uncertainties. The $\chi^2$ for the 4.39 $\mu$m and 15.2 $\mu$m bands are 2.8 and 3.0, respectively. It is worth noting that the contribution of these components remains the same when the CO$_2$:H$_2$O 1:1 100 K spectrum is exchanged for the 1:10 mixture at 10 K.

Finally, Figure \ref{fig:an-IRAS20126} shows that the long-wavelength wing of the bending mode is poorly fitted above 15.8 $\mu$m. In particular, there are two notable features at 16.2 $\mu$m and 16.5 $\mu$m that we were unable to reproduce with the laboratory data. These could be absorption features of crystalline silicates such as fosterite and enstatite. Additional evidence of these crystalline silicates was found at longer wavelengths with detections of the characteristic absorption band of crystalline fosterite at 23.2 $\mu$m \citep{hennign2010} but these features could also be present at shorter wavelengths. Removing these bands however requires a careful modeling of all the silicates features spanning the the 9 - 27 $\mu$m spectral region and this is beyond the scope of this paper.

\subsection{Low mass protostar: Per-emb 35}

After successfully fitting the two absorption features of IRAS 20126, we applied the linear combination to the ice absorption bands of the low-mass protostar Per-emb 35. This source is particularly interesting to study because the $^{12}$CO$_2$ combination mode located at 2.70 $\mu$m is observed with the NIRSpec G235H mode. {The three CO$_2$ vibrational modes in Per-emb 35 are successfully fitted with the same linear combination, the results are presented in Figure \ref{fig:an-per35}. The contribution of each individual component with respect to the total absorption is presented in Table \ref{Tab:per35}. These fractions are consistent between the three bands within the reported uncertainties and are also similar to the fractions derived in IRAS 20126. The alternative analysis with cold CO$_2$:H$_2$O ice is shown in Supplementary Figure \ref{fig:an-per35-cold}.

\begin{figure}[h!]
    \centering
     \includegraphics[width=1\hsize]{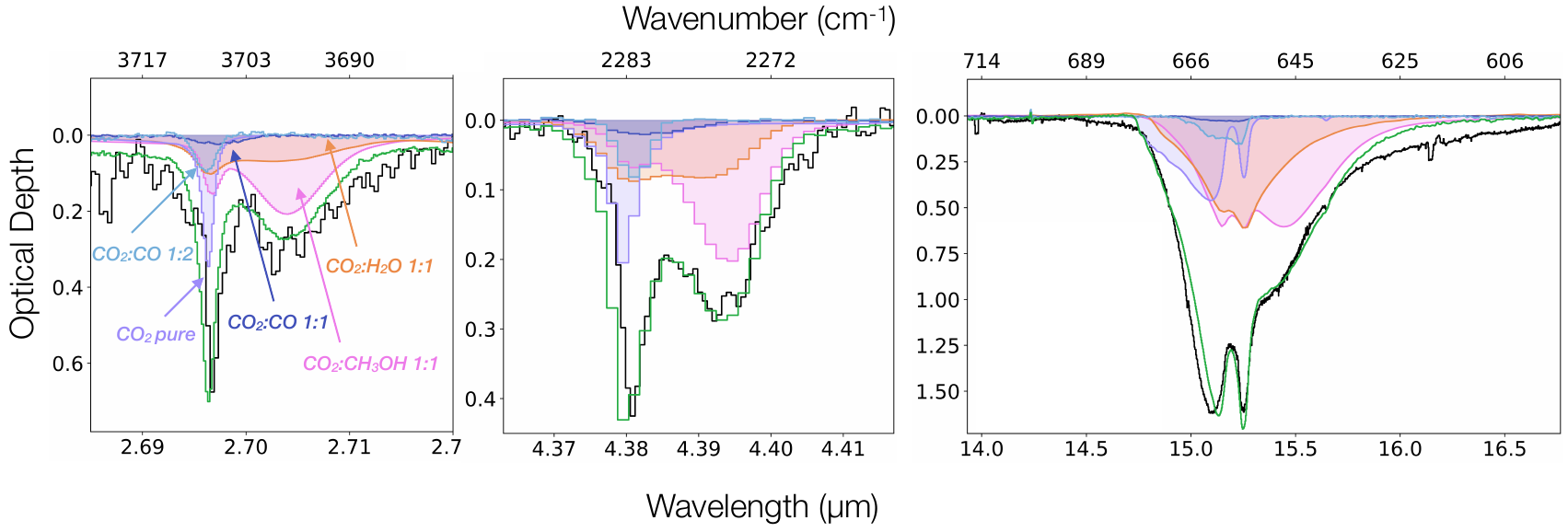}
    \caption{Band profile analysis of Per-emb 35. Left: Decomposition of the 2.70 $\mu$m $^{12}$CO$_2$ band. Middle: Decomposition of the 4.39 $\mu$m $^{13}$CO$_2$ band. Right: Decomposition of the 15.2 $\mu$m $^{12}$CO$_2$ band. The purple, pink, orange, light blue and dark blue shaded areas correspond to the pure CO$_2$ 80 K, CO$_2$:CH$_3$OH 1:1 115 K, CO$_2$:H$_2$O 1:1 100 K, CO$_2$:CO 1:2 25 K and CO$_2$:CO 1:1 15 K component, respectively. Finally the green line shows the sum of all the components. }
    \label{fig:an-per35}
\end{figure}

The sharp short-wavelength feature observed at 4.38 $\mu$m is successfully fitted with the laboratory spectrum of pure CO$_2$ ice at 80 K. The pure CO$_2$ 80 K spectrum also reproduces the short-wavelength peak observed at 2.69 $\mu$m in the $^{12}$CO$_2$ combination mode as well as the split peak profile in the bending mode at 15.2 $\mu$m. This pure CO$_2$ component is typically $\sim$ 15 - 20\% of the total integrated optical depth in both IRAS 20126 and Per-emb 35 and the strong contribution of this component indicates that these ices are undergoing segregation. The $\chi^2$ value for the fittings of the 2.70 $\mu$m, 4.39 $\mu$m and 15.2 $\mu$m bands are 7.54,  0.2 and 5.9, respectively.

\begin{center}
\begin{table*}[hbt!]
\caption{Fraction of integrated optical depth Per-emb 35}
\small
\centering
\begin{tabular}{lcccc}
\hline \hline
Component & mixture & 2.70 $\mu$m (\%) & 4.39 $\mu$m  (\%)  & 15.2 $\mu$m  (\%)
\\     
\hline 

CO$_2$:H$_2$O & 1:10 & 28 $\pm$ 4 & 28 $\pm$ 9  & 31 $\pm$ 1 \\
CO$_2$:CH$_3$OH & 1:1 & 50 $\pm$ 3 & 44 $\pm$ 1 & 47 $\pm$ 1\\
CO$_2$:CO& 1:1 & 3 $\pm$ 3 & 4 $^{+6}_{-4}$ & 	1 $\pm$ 1 \\
CO$_2$:CO &1:2 & 6 $\pm$ 6 & 6  $\pm$ 2 &	4 $\pm$ 1 \\
CO$_2$ & Pure	& 14  $\pm$ 3 &	19 $\pm$ 7 &	17 $\pm$ 1\\

\hline
\end{tabular}
\label{Tab:per35}
\end{table*}  
\end{center}

We note that the contribution of CO$_2$-CO mixed ices, previously inferred to be present for cold sources \citep{pontoppidan2008c2d}, is very small in both IRAS 20126 and Per-emb 35. This lack of CO ice is however consistent with the weak $^{12}$CO ice absorption band observed at 4.67 $\mu$m in both sources and the strong rotation-vibrational lines of gaseous CO seen in the 4 $\mu$m region of their spectra (Figure \ref{fig:full-per35}). In cold sources the 4.67 $^{12}$CO $\mu$m ice band is strong and often saturated ($\tau$ > 6) and there is usually no strong OCN$^-$ ice feature at 4.60 $\mu$m \citep{pontoppidan2003m, brunken2024b}. In contrast, the 4.67 $\mu$m CO ice band in IRAS 20126 has a peak optical depth of $\tau$ $\sim$ 1.3. This is relatively small compared to the strong OCN$^-$ feature observed in this source ($\tau$ $\sim$ 1.45) \citep{rubinstein2023}. Similarly, the 4.67 $\mu$m band in Per-emb 35 has a peak optical depth of $\tau$ $\sim$ 1.3 and has a strong OCN$^-$ band of $\tau$ $\sim$ 0.7 (Figure \ref{fig:full-per35}). These spectral features all point towards thermal desorption of the bulk of the CO ice. The CO$_2$-CO components are also more degenerate due to their complete overlap with the other components and as a result they also have the largest relative uncertainties.

\subsection{The bending mode of $^{13}$CO$_2$ ice}

A weak feature is observed at 15.64 $\mu$m overlaid on the long-wavelength wing of the bending mode. We detect this feature in both IRAS 20126 and Per-emb 35 (Figure \ref{fig:13co2-bend}). The small absorption band is also observed in the laboratory spectrum of pure CO$_2$ leading us to conclude that this feature is likely the bending mode of $^{13}$CO$_2$.

\begin{figure}[h!]
    \centering
     \includegraphics[width=1\hsize]{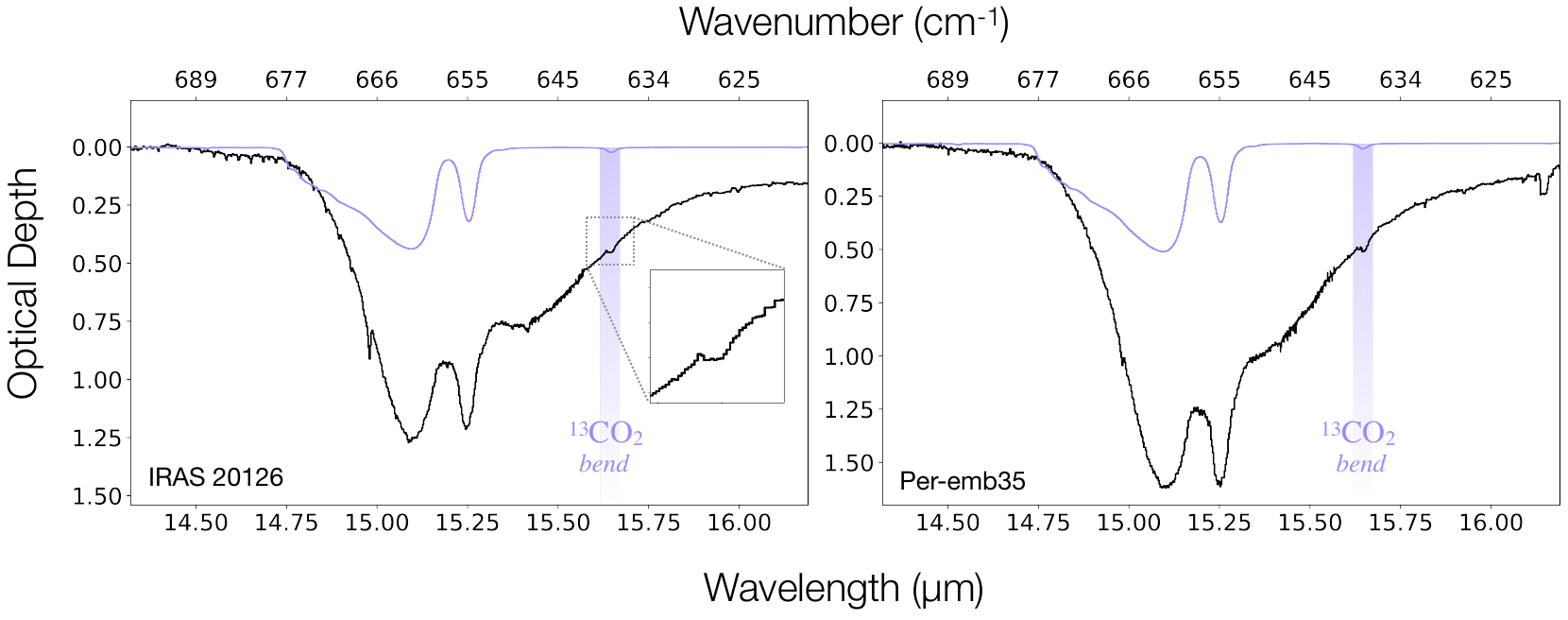}
    \caption{Bending mode of $^{13}$CO$_2$ ice. The 15.2 $\mu$m bending of $^{12}$CO$_2$ is shown on optical depth scale (black) and the purple shaded column shows the absorption feature overlaid on the wing of the band at 15.64 $\mu$m. The purple line shows the laboratory spectrum of pure CO$_2$ at 80 K with a small absorption feature centered at 15.64 $\mu$m.}
    \label{fig:13co2-bend}
\end{figure}

This band is only clearly visible in the spectrum of pure CO$_2$ ice likely because it produces sharp narrow features as can be seen at 2.69 $\mu$m, 4.38 $\mu$m and 15.2 $\mu$m. When CO$_2$ is diluted in other ices such as H$_2$O and CH$_3$OH this band probably becomes shallower and broader and therefore more difficult to detect. Because the $^{13}$CO$_2$ bending mode feature is small and there is a large uncertainty on the continuum needed to isolate it from the main $^{12}$CO$_2$ ice band, we refrain from doing further quantitative analysis on this feature in this work.


\subsection{$^{12}$C/$^{13}$C}

We quantified $^{12}$CO$_2$ and $^{13}$CO$_2$ column densities from the 15.2 $\mu$m and 4.39 $\mu$m features, respectively, and derived $^{12}$C/$^{13}$C = 90 $\pm$ 9 in IRAS 20126. The $^{12}$CO$_2$ and $^{13}$CO$_2$ column densities in Per-emb 35 were measured from the 2.70 $\mu$m band, the 4.39 $\mu$m band and the 15.2 $\mu$m by \citet{brunken2024b} and the $^{12}$C/$^{13}$C ratio derived for the 2.70 $\mu$m combination mode and the 15.2 $\mu$m bending mode are 132 and 99, respectively.

The observational errors are small, $\sim$ 10\% and the error uncertainty due to the continuum are up to $\sim$ 20\%. The majority of the error budget, not included in the above error margins, comes from the uncertainties on the band strengths and can be up to $\sim$ 46\%. For further details on the error analysis we refer the reader to \citet{brunken2024b}. The findings are summarized in Table \ref{Tab:column}.

The $^{12}$C/$^{13}$C$_{ice}$ = 90 $\pm$ 9 measured in IRAS 20126 is consistent with the ratios measured from the CO$_2$ vibrational modes by \citet{brunken2024b} in the envelopes of low-mass protostars (their Table \ref{Tab:column}, mean ratio measured from the 15.2 $\mu$m bending modes $\sim$ 97 $\pm$ 17). It is slightly elevated with respect to the ratio measured for the ISM $\sim$ 68 \citep{boogert1999iso}. This value is at the lower end of the  gas-phase ratio previously measured in this same source by \citet{rubinstein2023} from hot gaseous CO rotation-vibrational lines  $^{12}$C/$^{13}$C$_{gas}$ > 106.

\begin{center}
\begin{table}[hbt!]
\caption{$^{12}$CO$_2$ and $^{13}$CO$_2$ ice column densities in cm$^{-2}$.}
\small
\centering

\begin{tabular}{cccccc}
\hline \hline
Source & N $^{12}$CO$_2$ & N $^{13}$CO$_2$ & N $^{12}$CO$_2$  &  $^{12}$C/$^{13}$C & $^{12}$C/$^{13}$C \\
& 2.70 $\mu$m & 4.39 $\mu$m  & 15.2 $\mu$m &   2.70 $\mu$m &  15.2 $\mu$m
\\     
\hline 

IRAS 20126 & - & 2.3 $\times$ 10$^{16}$ & 2.1 $\times$ 10$^{18}$  & - & 90 $\pm$ 9 \\
Per-emb 35 & 	3.7  $\times$ 10$^{18}$ & 2.8  $\times$ 10$^{16}$ &	2.8  $\times$ 10$^{18}$  & 132 $\pm$ 13 & 99 $\pm$ 10 \\

\hline
\end{tabular}
\label{Tab:column}
\end{table}  
\end{center}



\section{Discussion}
\label{sec:discussion}

\subsection{Formation and segregation of CO$_2$ ice}

CO$_2$ chemistry occurs during both the H$_2$O-dominated phase and the CO-dominated phase of interstellar ice formation \citep{boogert1999iso}. The first formation route is facilitated by the formation of the HO-CO complex through the radical reaction\citep{oba2010experimental,ioppolo2011surface,oberg2011spitzer}:

\begin{equation}
    \ch{CO + OH -> HO-CO -> CO_2 + H } \label{eq:pathway1}
.\end{equation}

The CO$_2$-H$_2$O ices that ensue from this reaction are observed in all the vibrational modes of CO$_2$ \citep{gerakines1999infrared, boogert1999iso,pontoppidan2008c2d,brunken2024a,brunken2024b} and are part of the polar H$_2$O-rich ice layer that forms during the during the translucent cloud phase when atomic H is abundant.

Apolar CO$_2$-CO ices subsequently encapsulate the water-rich ice layer during the catastrophic CO freeze-out epoch \citep{pontoppidan2006} when OH reacts with CO on the cold grains. The high densities and low temperatures during this stage make it possible for CO to freeze out in large amounts on the dust grains spawning these CO-rich apolar ices. Additionally, the freeze-out also triggers the formation of molecules that use CO as feedstock such as methanol (CH$_3$OH), another key ingredient in interstellar ices \citep{allamandola1992} and a molecule that is known to be intimately mixed with CO$_2$ \citep{ehrenfreund1998ice,ehrenfreund1999laboratory, gerakines1999infrared,boogert1999iso,pontoppidan2008c2d,brunken2024a,brunken2024b}. 

\begin{figure}[h!]
    \centering
     \includegraphics[width=1\hsize]{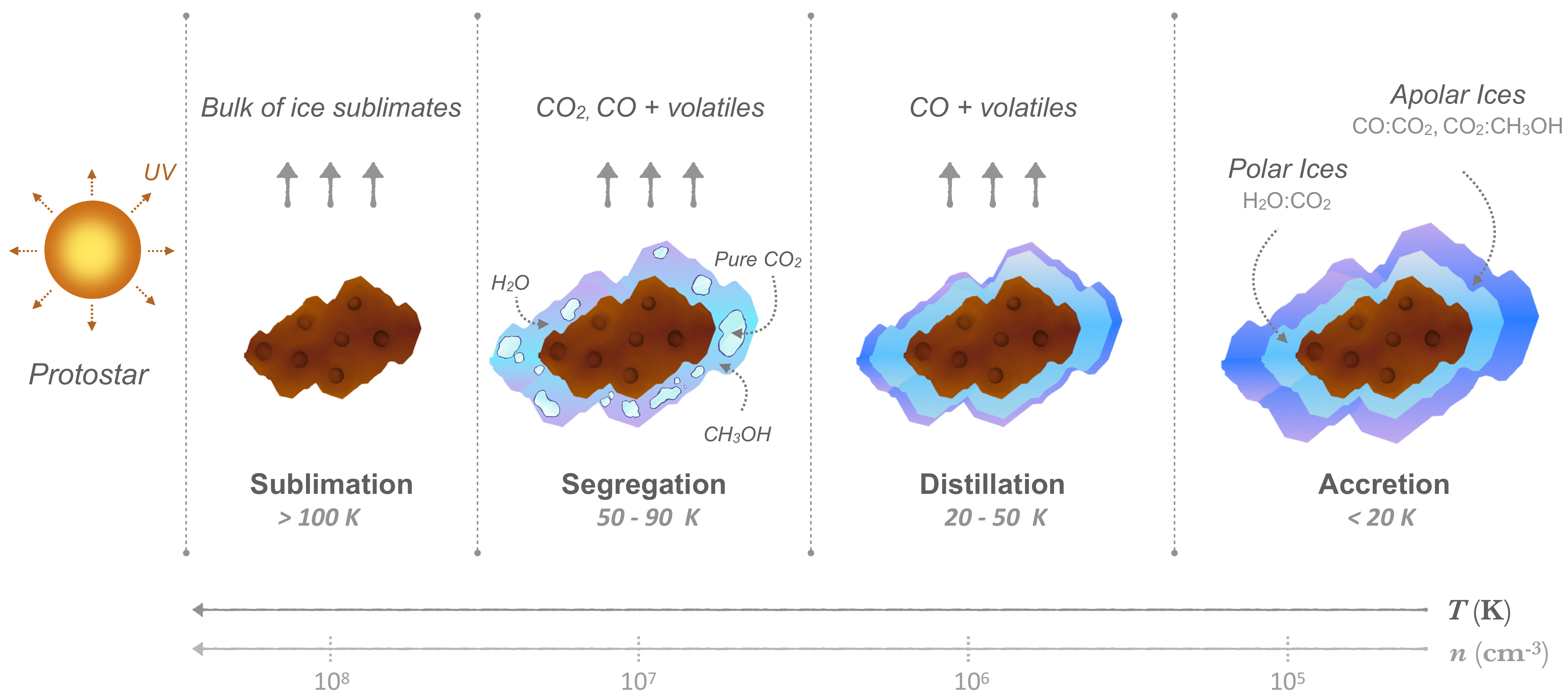}
    \caption{Schematic overview of CO$_2$ ice segregation in protostellar envelopes. The temperature and density zones are separated by the gray dashed lines and each grain represents a different stage of ice processing. Observations probe the entire line of sight from the central regions around the protostar to the outer envelope.  }
    \label{fig:segregation}
\end{figure}

When the young protostar heats its surrounding envelope, the most volatile species, including CO ice, sublimate from the grains \citep{oberg2011spitzer,pontoppidan2008c2d} as illustrated in Figure \ref{fig:segregation}. This CO desorption is evident in IRAS 20126 and Per-emb 35 from the weak CO ice band at 4.67 $\mu$m and the strong gaseous CO line-forest spanning the 4 $\mu$m spectral region \citep{brunken2024a,brunken2024b}. This distillation process is also consistent with the experimental results from \citet{ehrenfreund1998ice} who showed that above $\sim$ 50 K H$_2$O, CO$_2$ and CH$_3$OH become the dominant species in interstellar ice spectra.

After distillation, segregation of CO$_2$ quickly follows as temperatures continue to rise in the envelope. During this segregation, the water molecules will rearrange and form links through hydrogen bonds. The new spatial proximity will allow the CO$_2$ molecules to cluster together and form bonds, producing inclusions of pure CO$_2$ ice inside the ice mantle \citep{ehrenfreund1999laboratory} (Figure \ref{fig:segregation}). Because water desorbs at a much higher temperature than CO$_2$, the pure CO$_2$ ice will remain trapped on the grain well beyond its desorption temperature. This multilayer formation mechanism coupled with the physicochemical processes that follow after, produce the different components that constitute the ice absorption bands observed in protostellar spectra. These absorption features include not only the components that are present on individual dust grains, but also the absorption of all the ices along the line sight. This extends from the hot central regions, to the ices in the colder outer envelope \citep{ehrenfreund1998ice} (Figure \ref{fig:segregation}). Envelopes that are exposed to higher levels of protostellar heating, such as those around luminous protostars, have higher overall temperature and thus experience higher degrees of ice processing.

The segregation of pure CO$_2$ ice is corroborated by the profile analyses done on the absorption bands in IRAS 20126 and Per-emb 35. Figures \ref{fig:an-IRAS20126} and \ref{fig:an-per35} show that a pure CO$_2$ component is observed in all the vibrational modes of CO$_2$ that accounts for up to $\sim$ 20\% of the total absorption band. At 2.69 $\mu$m and 4.38 $\mu$m this pure CO$_2$ ice component creates a narrow secondary absorption feature on the short-wavelength side of the bands. At 15.2 $\mu$m pure CO$_2$ ice produces the characteristic double peak structure. These features are not detected in the spectra of non-heated sources \citep{brunken2024a,brunken2024b}.

Another intrinsic feature of ice segregation is the deep shoulder observed at 15.4 $\mu$m that is associated with CO$_2$-CH$_3$OH ices \citep{ehrenfreund1998ice,dartois1999}. This shoulder is caused by the acid-base interaction between the carbon atom of CO$_2$ and the hydrogen atom of CH$_3$OH. \citet{ehrenfreund1998ice} and \citet{dartois1999} both demonstrated with experimental studies that segregation is vital for the appearance of this shoulder which becomes observable in laboratory spectra with temperatures ranging between 65 K - 110 K. Many astronomical observations also support these findings, with numerous detections of this feature in high-mass and luminous protostars \citep{dartois1999} including several massive young stellar objects populating the Central Molecular Zone (CMZ) ($R_{gal}$ < 200 pc) \citep{An2011}. 

This shoulder is well fitted with a laboratory mixture of CO$_2$:CH$_3$OH 1:1 ice at 115 K (Figures \ref{fig:an-IRAS20126} and \ref{fig:an-per35}). This is consistent with the temperature range reported in \citet{dartois1999} and the findings in \citet{ehrenfreund1998ice} who showed that astronomical data are better fitted with 1:1 laboratory mixtures. The CO$_2$:CH$_3$OH 1:1 at 115 K spectrum also successfully reproduces the long-wavelength absorption features observed at 2.70 $\mu$m and 4.39 $\mu$m.  

The values presented in Tables \ref{Tab:IRAS20126} and \ref{Tab:per35} show that fractions of integrated optical depths are similar within reported error margins. From these results it can inferred that the $^{12}$C/$^{13}$C ratios are also similar among the different ice components. This indicates that the carbon isotope ratio is likely set at the formation stage of the different CO$_2$ ices and that fractionation processes such as isotope exchange reactions \citep{watson1976, langer1984, langerpenzias1993} appear to not have played a significant role during neither the water-dominated phase nor the heavy CO freeze-out stage that produce the CO$_2$:H$_2$O and CO$_2$:CH$_3$OH ices. This ratio persists throughout the heating stage that produces the pure CO$_2$ component. We note however that this conclusion is drawn solely on the similarity between these fractions and that the isotope ratios have not been calculated for these individual components due to lack of experimental data. To the best of our knowledge there are currently no band strengths for CO$_2$:CH$_3$OH mixtures for instance and band strengths measured at different temperatures are also limited. Quantifying the carbon isotope ratio for these individual components will therefore result in large uncertainties \citep{brunken2024b}.

Tables \ref{Tab:IRAS20126} and \ref{Tab:per35} and Figures \ref{fig:an-IRAS20126} and \ref{fig:an-per35} show that CO$_2$:CH$_3$OH ices are the main contributors of the CO$_2$ vibrational modes. This conclusion differs from the findings presented in \citet{pontoppidan2008c2d} that fitted the 15.2 $\mu$m bands with mostly with CO$_2$-H$_2$O ices. Our results also differ from the analysis presented in \citet{slavicinska2024} who fitted the 3.5 $\mu$m CH$_3$OH band in IRAS 20126 with mostly CH$_3$OH:H$_2$O and pure CH$_3$OH ices and found no contribution of CO$_2$:CH$_3$OH ices. These differences are discussed in the following section.

\subsection{CO$_2$-CH$_3$OH vs. CO$_2$-H$_2$O ices}

In contrast to the analysis presented in \citet{pontoppidan2008c2d} our results suggest that the CO$_2$:CH$_3$OH component accounts for more than 40\% of the three CO$_2$ vibrational modes in both IRAS 20126 and Per-emb 35. To understand the source of this difference, we revisited the band profile analysis and fitted the 15.2 $\mu$m absorption features of IRAS 20126 and Per-emb 35 with the components used in \citet{pontoppidan2008c2d}. The results indicate that the water component remains dominant in this particular linear combination because \citet{pontoppidan2008c2d} simulated the 15.4 $\mu$m shoulder with a combination of two Gaussian curves. However, these Gaussian profiles only model the long-wavelength component of the CO$_2$:CH$_3$OH ices centered at 15.4 $\mu$m. The short wavelength component located at 15.2 $\mu$m, which appears in all laboratory spectra of CO$_2$:CH$_3$OH ices \citep{ehrenfreund1999laboratory}, are not included in this model. As a result, the isolated long-wavelength feature creates a sharp shoulder that remains detectable in the spectrum even when the contribution of the  CO$_2$:H$_2$O component greatly exceeds that of the CO$_2$:CH$_3$OH component.  

In contrast, when laboratory spectra of CO$_2$:CH$_3$OH ices, containing both long- and short-wavelength components, are used, this shoulder becomes shallower in the final linear combination. This in turn will require a larger contribution from the CO$_2$:CH$_3$OH ice to reproduce the shoulder and a smaller contribution from the CO$_2$:H$_2$O component to avoid overfitting the band. Moreover, the short-wavelength component in the CO$_2$:CH$_3$OH laboratory spectra also constrains the contribution of the pure CO$_2$ component at 15.2 $\mu$m since a balance between the two components is needed to avoid overproducing the double peak feature. 

To confirm that this large fraction of CO$_2$:CH$_3$OH ices is indeed caused by segregation, the ice absorption bands of two cold sources in the JOYS+ sample, Per-emb55-a and EDJ183-a, were also fitted. The results showed that the CO$_2$ ice bands in these cold sources have inherently larger fractions of CO$_2$:H$_2$O ices. The band profile analysis presented in \citet{brunken2024a} also shows a large contribution of CO$_2$:H$_2$O ices in the $^{13}$CO$_2$ bands of cold protostars. The small contribution of this component in our heated protostars therefore indicates that segregation must be occurring for the most part in the H$_2$O-rich ice layer. If so, then these CO$_2$:H$_2$O ices should be both heated and showing signs of ice segregation. This is supported by the analysis where CO$_2$:H$_2$O 1:1 spectra at 100 K are used to fit CO$_2$ ice features. Not only are there clear signs of ice segregation in the 1:1 mixture between 90 - 120 K with the appearance of the pure CO$_2$ peak at 4.38 $\mu$m (Figure \ref{fig:boogert}), but this temperature range is also consistent with the temperature range of the CO$_2$:CH$_3$OH component (115 K). Experimental results by \citet{he2018} also show that CO$_2$ needs to account for at least 23\% of the ice mixture for segregation to take place, making the CO$_2$:H$_2$O 1:1 spectrum a more appropriate choice for fitting the ice bands than the 1:10 mixture. While we do expect to have some contribution of cold ices in the outer envelope, the spectral features in both IRAS 202126 and Per-emb 35 indicate that the majority of the ices must be heated. \citet{slavicinska2024} for instance showed that the 3 $\mu$m H$_2$O band of IRAS 20126 has a crystalline profile. We reiterate that exchanging the cold CO$_2$:H$_2$O spectrum for the warm CO$_2$:H$_2$O spectrum does not change the fractions of integrated optical depths. An intriguing question for future investigation is whether the total abundance of CO$_2$:H$_2$O ices in cold sources equals the sum of the segregated pure CO$_2$ ice and the remaining CO$_2$:H$_2$O ices in heated sources. 


Regarding the CH$_3$OH 3.5 $\mu$m band in IRAS 20126, where \citet{slavicinska2024} showed that the feature is a composite of mostly pure CH$_3$OH ice and CH$_3$OH:H$_2$O ices, the differences could be due to different ways of fitting and subtracting the continuum over the 3.5 $\mu$m CH$_3$OH band. This weak band is overlaid on the wing of the 3 $\mu$m water band whose shape is highly susceptible to grain shape and size effects. This introduces an uncertainty on the actual shape of the continuum. \citet{dartois2001} for instance showed that the wing of the water band can be significantly raised with respect to the absorption feature at 3.5 $\mu$m. To test this, we revisited this band and raised the continuum over this region as shown in the left panel of Figure \ref{fig:continuum-ch3oh}. The results indicate that CO$_2$:CH$_3$OH can contribute significantly to this band if we use this approach (Figure \ref{fig:continuum-ch3oh}, right panel). Additionally, prior to the analysis, PAH features peaking in this spectral region had to be removed which further adds to the uncertainties on the 3.5 $\mu$m feature. For the PAH removal, we followed the methods described \citet{slavicinska2024}. Finally, we note that the 3.5 $\mu$m band overlaps with absorption features of ammonia hydrates that also peak at these wavelengths \citep{boogert2022}.

\section{Conclusions}
\label{sec:conclusions}

We present band profile analyses of the $^{12}$CO$_2$ 15.2 $\mu$m bending mode, the $^{13}$CO$_2$ 4.39 $\mu$m stretching mode and the $^{12}$CO$_2$ 2.70 $\mu$m combination mode that use a consistent set of laboratory ice mixtures in the high mass protostar IRAS 20126 and the low mass protostar Per-emb 35. The spectra of both sources show clear signs that the ices in the envelope are being heated by the central protostar prompting segregation of pure CO$_2$ ice.

\begin{itemize}
  \item The 15.2 $\mu$m double peak feature and the short-wavelength features at 2.69 $\mu$m and 4.38 $\mu$m are successfully fitted with laboratory spectra of pure CO$_2$ ice at 80 K pointing towards segregation due to protostellar heating. This pure CO$_2$ component accounts for $\sim$ 20\% of the total absorption and is consistent across all the vibrational modes and between the low-mass and the high mass-protostar. 
  
  \item The deep shoulder observed at 15.4 $\mu$m is reproduced the laboratory spectrum of CO$_2$:CH$_3$OH 1:1 ices at 115 K. This spectrum also fitted the long-wavelength features at 2.70 $\mu$m in Per-emb 35 and 4.39 $\mu$m in both sources. This CO$_2$:CH$_3$OH component is dominant in all the vibrational modes and  accounts for more than 40\% of the total band absorption. This is contrary to what has been previously observed in the CO$_2$ bands of non-heated sources where the CO$_2$:H$_2$O ices are the main contributor. This indicates that CO$_2$ is likely segregating from mostly the water-rich ice layer.
  
  \item The contribution of each component with respect to the total integrated optical depth is similar in all the vibrational modes. This suggests that the $^{12}$C/$^{13}$C ratio is already set at the formation stage of the different CO$_2$ ices and that fractionation processes did not play a significant role. 
  
  \item We report the detection of the bending mode of $^{13}$CO$_2$ ice at 15.64 $\mu$m. The feature is overlaid on the long-wavelength wing of the $^{12}$CO$_2$ 15.2 $\mu$m band.
  
  \item We quantified the column densities and derived a $^{12}$C/$^{13}$C$_{ice}$ $\sim$ 90 in IRAS 20216. This value is slightly higher compared to the value measured for the ISM and lower than the gas-phase ratio recently derived for IRAS 20126. The $^{12}$C/$^{13}$C$_{ice}$ of Per-emb 35 was previously measured to be $\sim$ 99 and $\sim$ 132 for the 15.2 $\mu$m bending mode and the 2.70 $\mu$m combination mode, respectively, consistent with the values measured for other low mass sources. The observational uncertainties are $\sim$ 10\%.
 \end{itemize} 
The findings in this work indicate that segregation dramatically changes the structure of these ices and that this process occurs in a similar manner in both the high-mass and the low-mass protostar. Future work should investigate whether the total abundance of CO$_2$:H$_2$O ices in non-heated sources matches the sum of the segregate pure CO$_2$ ice and the remainder of the CO$_2$:H$_2$O ices in heated sources. Furthermore, more laboratory data of CO$_2$ band strengths are needed, in particular band strengths were CO$_2$ is mixed with other ices, such as CH$_3$OH, as well as band strengths measured at different temperatures. These measurements are crucial to accurately determine the column density of these ices and the carbon isotope ratio. Finally, since CO$_2$ and H$_2$O ice are originally intimately mixed, additional laboratory data with new more mixing ratios are needed to better study the chemical evolution of this component in both heated and cold sources.

\begin{acknowledgement}
The lead authors have had the pleasure of working closely with Prof. Harold Linnartz for many years on interstellar ices, combining his work in the Laboratory for Astrophysics with astronomical observations. The spectroscopy and chemistry of CO$_2$ ice has been a focus of his research in Leiden and this work is therefore a salute to his contributions to Astrochemistry. Astrochemistry in Leiden is supported by the Netherlands Research School for Astronomy (NOVA), by funding from the European Research Council (ERC) under the European Union’s Horizon 2020 research and innovation programme (grant agreement No. 101019751 MOLDISK. Support by the Danish National Research Foundation through the Center of Excellence “InterCat” (Grant agreement no.: DNRF150) is also acknowledged. This work is based on observations made with the NASA/ESA/CSA James Webb Space Telescope. The data were obtained from the Mikulski Archive for Space Telescopes at the Space Telescope Science Institute, which is operated by the Association of Universities for Research in Astronomy, Inc., under NASA contract NAS 503127 for JWST. The following National and International Funding Agencies funded and supported the MIRI development: NASA; ESA; Belgian Science Policy Office (BELSPO); Centre Nationale d’Etudes Spatiales (CNES); Danish National Space Centre; Deutsches Zentrum fur Luftund Raumfahrt (DLR); Enterprise Ireland; Ministerio De Economiá y Competividad; Netherlands Research School for Astronomy (NOVA); Netherlands Organisation for Scientific Research (NWO); Science and Technology Facilities Council; Swiss Space Office; Swedish National Space Agency; and UK Space Agency.



\end{acknowledgement}


\bibliography{main}

\begin{suppinfo}

Supporting information: Additional figures including continuum fittings in the 15 $\mu$m region, alternative band profile fittings of the CO$_2$ ice bands with different laboratory spectra, band profile fitting of the 3.5 $\mu$m methanol feature and tables containing properties of the two sources and the FWHM and central positions of the Gaussian curves fitted to the $^{13}$CO$_2$ ice band.

\begin{figure}[h!]
    \centering
     \includegraphics[width=1\hsize]{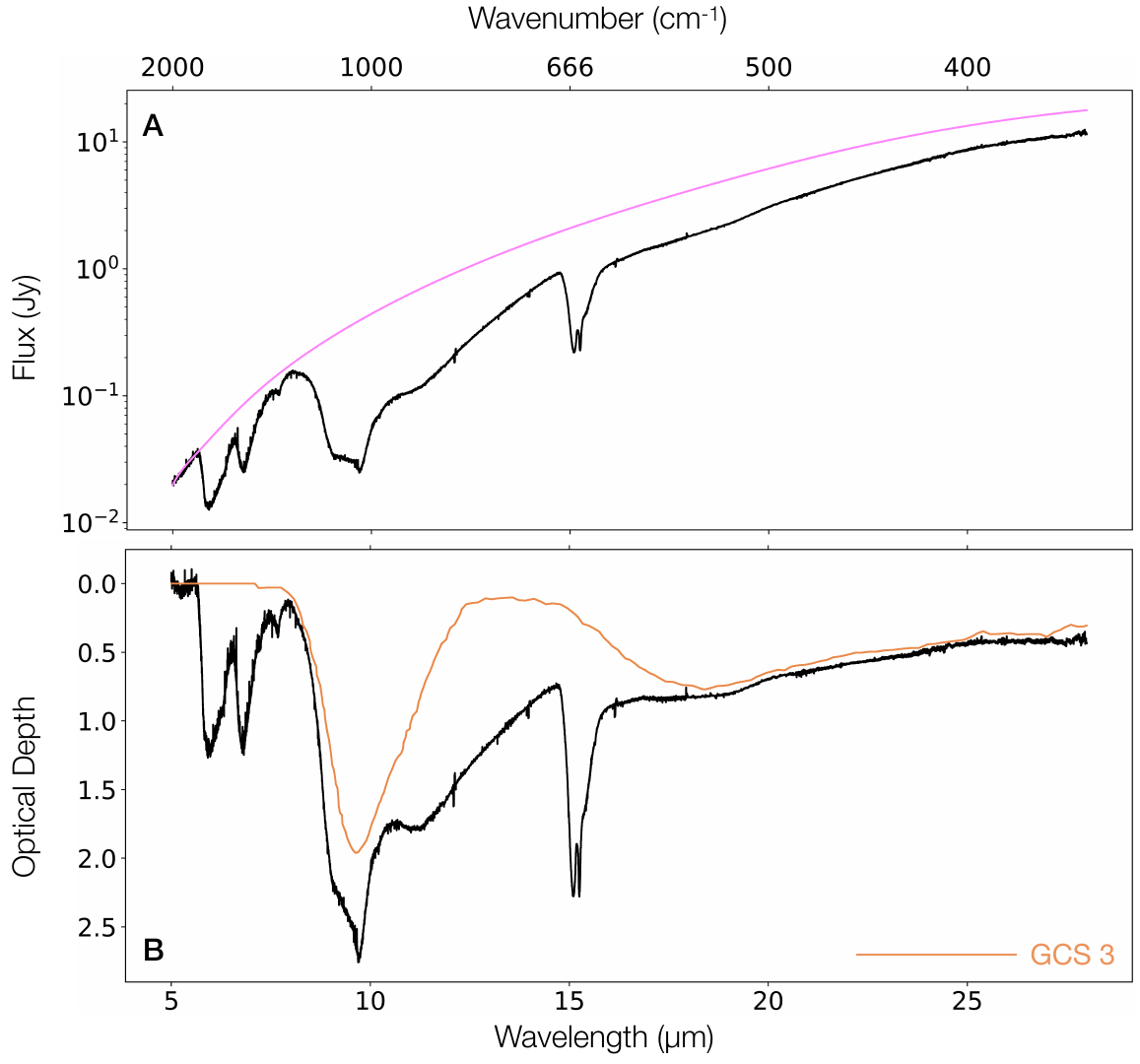}
    \caption{Continuum determination for Per-emb 35. Panel A shows the global continuum fitted over the 5 - 28 $\mu$m region (pink). Panel B shows the spectrum of GCS 3 fitted over the continuum subtracted spectrum (orange). }
    \label{fig:continuum-per35}
\end{figure}

\begin{figure}[h!]
    \centering
     \includegraphics[width=1\hsize]{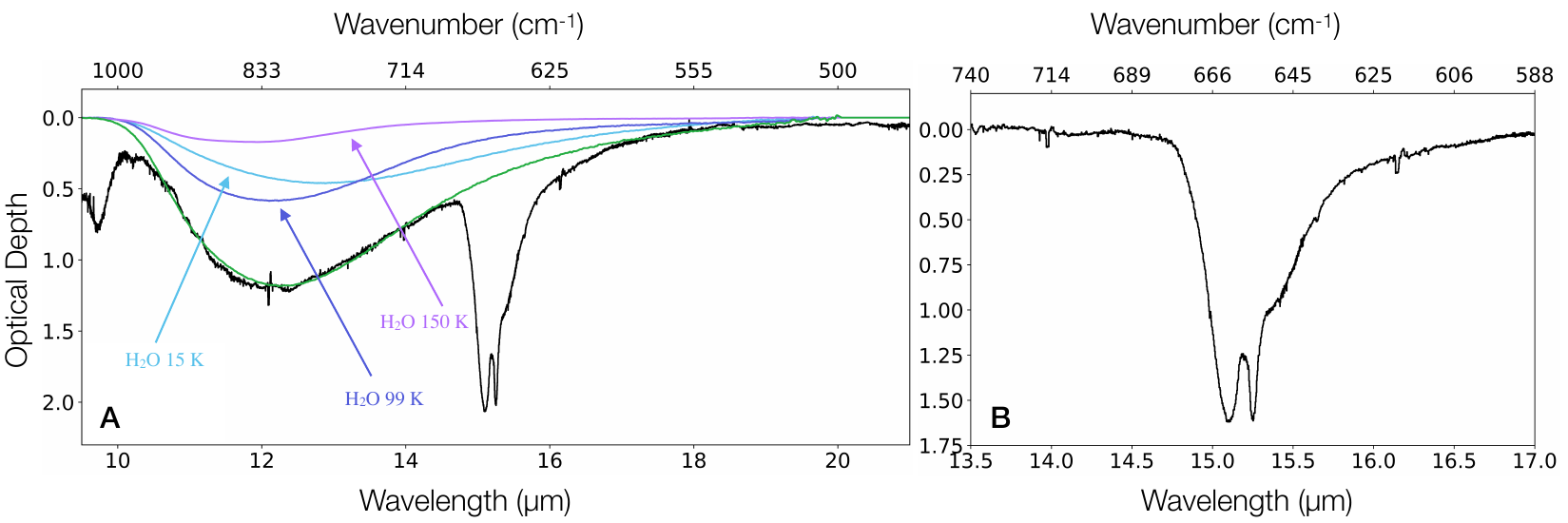}
    \caption{Subtraction of the water libration mode in Per-emb 35. Panel A shows the spectrum of water ice at different temperature fitted over the silicate subtracted spectrum. Panel B shows the final spectrum on optical depth scale.  }
    \label{fig:water-lib-per35}
\end{figure}

\begin{figure}[h!]
    \centering
     \includegraphics[width=1\hsize]{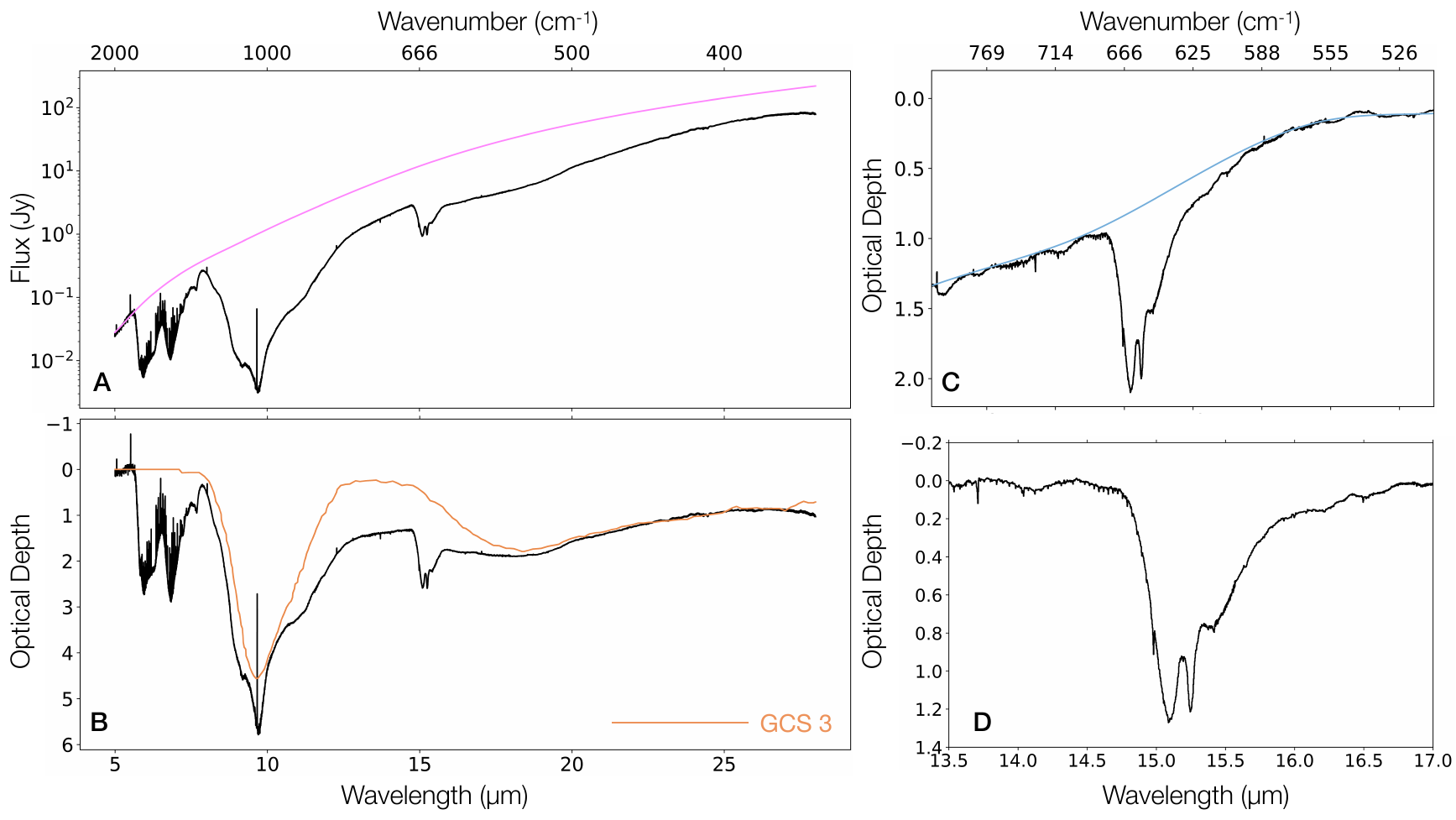}
    \caption{Continuum determination for IRAS 20126. Panel A shows the global continuum fitted over the 5 - 28 $\mu$m region (pink). Panel B shows the spectrum of GCS 3 fitted over the continuum subtracted spectrum (orange). Panel C shows the local continuum fitted over the silicate subtracted spectrum to simulate the wing of the water libration mode. Panel D shows the final spectrum on optical depth scale. }
    \label{fig:continuum-iras20126}
\end{figure}

\begin{center}
\begin{table}[hbt!]
\caption{Source properties.}
\small
\centering
\begin{tabular}{lcccl}
\hline \hline
Source & RA & Dec & L (L$_{\odot}$) & Distance (pc)   \\         
\hline 

IRAS 20126 & 20:14:26.04 & 41:13:32.43 & 10$^4$ & 1550 \\
Per 35 & 3:28:37.093  &  31:13:30.83 & 9.3 & 293 \\

\hline
\end{tabular}
\label{Tab:properties}
\end{table}  
\end{center}\

\begin{center}
\begin{table}[hbt!]
\caption{Properties of the spectral features comprising the $^{13}$CO$_2$ band of IRAS 20126.}
\small
\centering
\begin{tabular}{lccc}
\hline \hline
Component &  Peak Position (cm$^{-1}$)  & FWHM (cm$^{-1}$) 
\\     
\hline

Long-wavelength  & 2275.72 $\pm$ 0.1 & 6.6 $\pm$ 0.1 \\
Middle & 2278.90 $\pm$ 0.4 & 9.0 $\pm$ 0.6 \\
Short-wavelength & 2282.91 $\pm$ 0.01 & 3.5 $\pm$ 0.1 \\

\hline
\end{tabular}
\label{Tab:gaus}
\begin{tablenotes}\footnotesize
\item{} 
\end{tablenotes}
\end{table}  
\end{center}

\begin{figure}[h!]
    \centering
     \includegraphics[width=1\hsize]{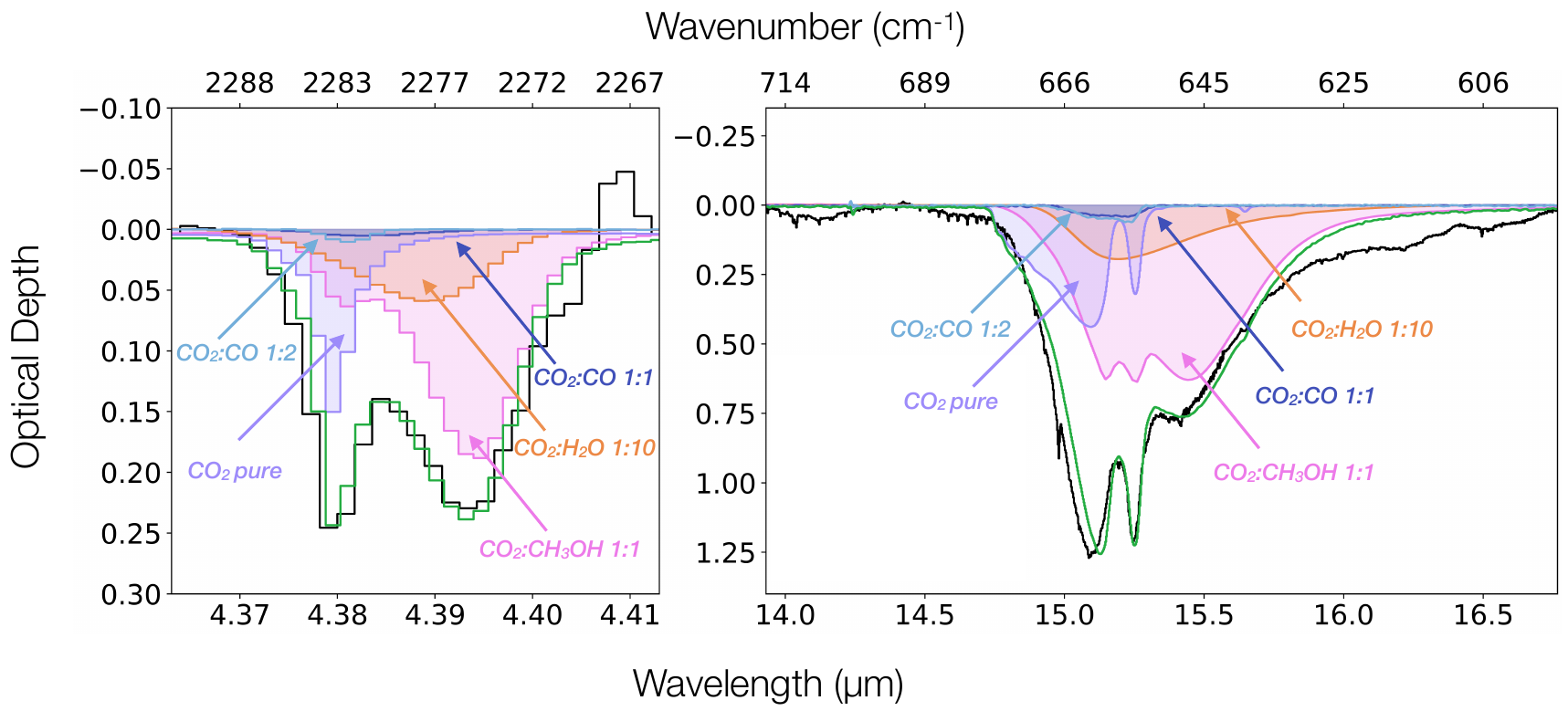}
    \caption{Alternative band profile analysis of IRAS 20126. Left: Decomposition of the 4.39 $\mu$m $^{13}$CO$_2$ band. Right: Decomposition of the 15.2 $\mu$m $^{12}$CO$_2$ band. The purple, pink, orange, light blue and dark blue shaded areas correspond to the pure CO$_2$ 80 K, CO$_2$:CH$_3$OH 1:1 115 K, CO$_2$:H$_2$O 1:10 10 K, CO$_2$:CO 1:2 25 K and CO$_2$:CO 1:1 15 K component, respectively. Finally the green line shows the sum of all the components. The poor fit at 16.2 $\mu$m and 16.5 $\mu$m is likely due to absorption features of crystalline silicates.}
    \label{fig:an-IRAS20126-cold}
\end{figure}

\begin{figure}[h!]
    \centering
     \includegraphics[width=1\hsize]{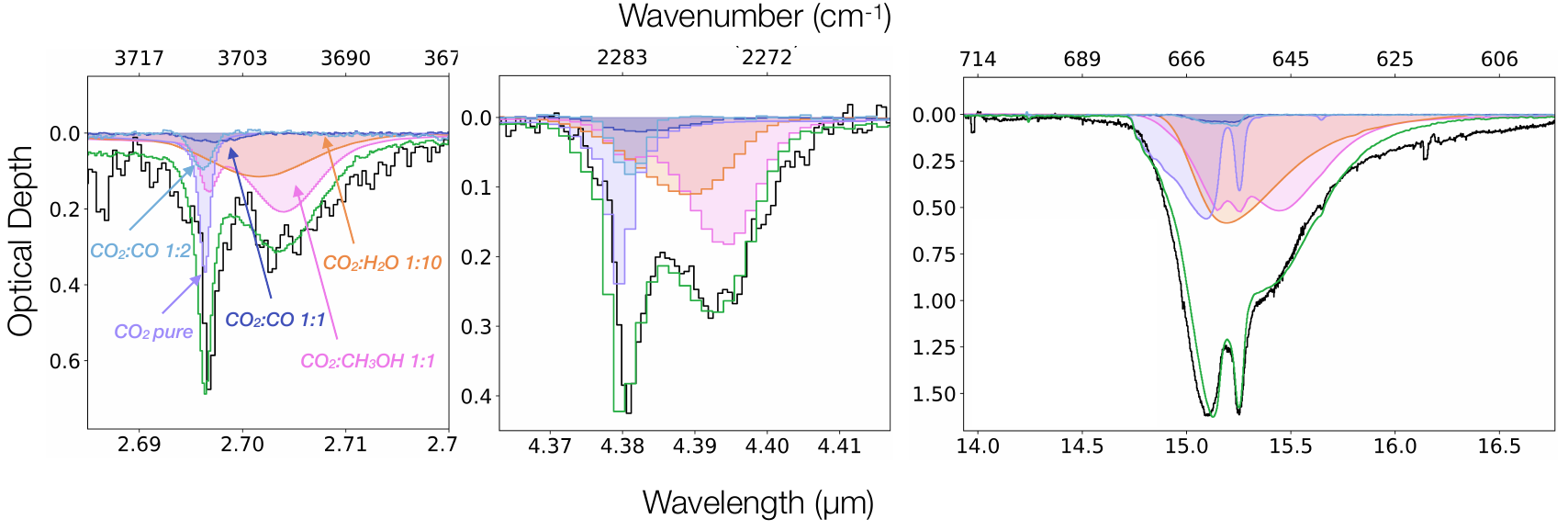}
    \caption{Alternative band profile analysis of Per-emb 35. Left: Decomposition of the 4.39 $\mu$m $^{13}$CO$_2$ band. Right: Decomposition of the 15.2 $\mu$m $^{12}$CO$_2$ band. The purple, pink, orange, light blue and dark blue shaded areas correspond to the pure CO$_2$ 80 K, CO$_2$:CH$_3$OH 1:1 115 K, CO$_2$:H$_2$O 1:10 10 K, CO$_2$:CO 1:2 25 K and CO$_2$:CO 1:1 15 K component, respectively. Finally the green line shows the sum of all the components. The poor fit at 16.2 $\mu$m and 16.5 $\mu$m is likely due to absorption features of crystalline silicates.}
    \label{fig:an-per35-cold}
\end{figure}

\begin{figure}[h!]
    \centering
     \includegraphics[width=1\hsize]{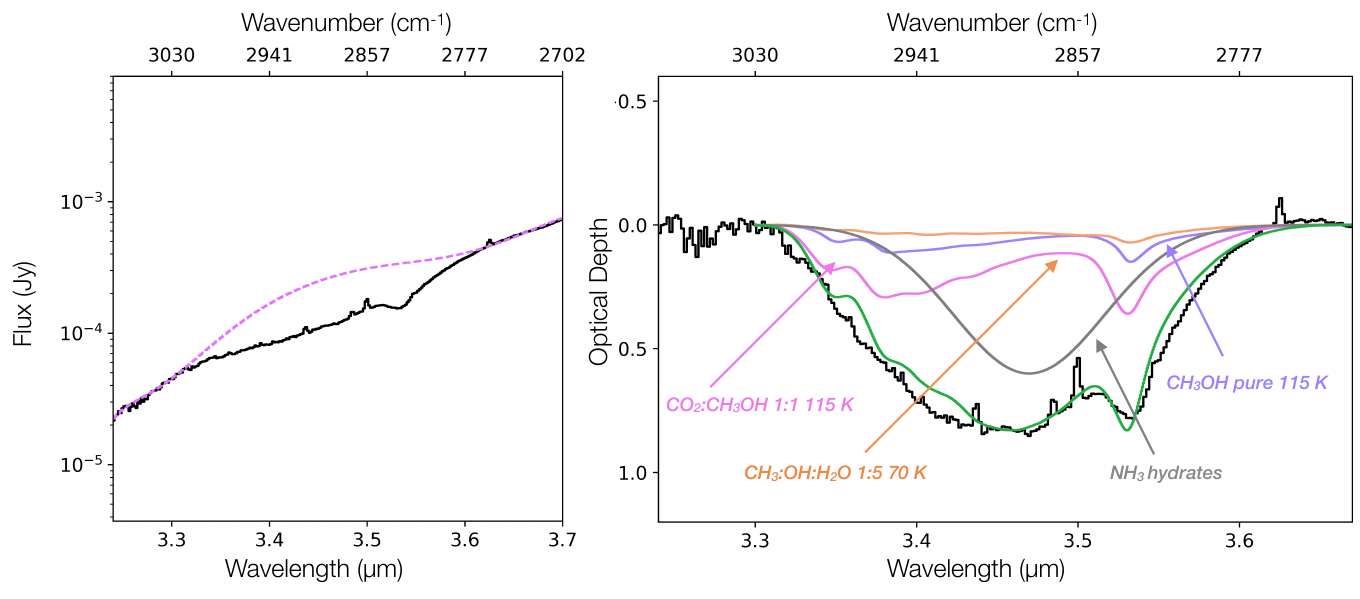}
    \caption{Continuum determination (left) and spectral decomposition (right) of the 3.5 CH$_3$OH feature. The purple, pink, orange, and gray lines correspond to spectra of pure CH$_3$OH 115 K, CO$_2$:CH$_3$OH 1:1 115 K, H$_2$O:CH$_3$OH 5:1 70 K and a Gaussian representing the ammonia hydrates in this region, respectively. The H$_2$O:CH$_3$OH and pure CH$_3$OH spectra were obtained from \citet{slavicinska2024} and the properties of the Gaussian curve are based on the results presented in \citet{boogert2022}.}
    \label{fig:continuum-ch3oh}
\end{figure}

\end{suppinfo}

\end{document}